\documentclass[12pt,a4paper]{article}   
\usepackage{amsmath,amsfonts,amssymb}

\newtheorem{theo}{Theorem}
\newtheorem{lemm}{Lemma}

\begin{document}

\title{ Characters for $\widehat{\mathfrak{sl}(n)}_{k=1}$  from
                    a novel Thermodynamic Bethe Ansatz\thanks{
      A part of the subject was presented at "International Workshops
      on Statistical Mechanics and Integrable Models",
      20 July-8 Aug. 1997 (Canberra)}    }
\author{
        J. Suzuki\thanks{e-mail: suz@hep1.c.u-tokyo.ac.jp}\\
        \parbox{0.9\textwidth}{
        {\em
        \begin{center}
		Institute of Physics,\\
        University of Tokyo at Komaba \\                
        Komaba 3-8-1, Meguro-ku, \\     
        Tokyo, JAPAN  \\                   
        \end{center}
        }}  
       }    

\date{January 1998}

\maketitle
\begin{abstract}
Motivated by the recent development on the exact
thermodynamics of 1D quantum systems, we propose
 quasi-particle like formulas for $\widehat{\mathfrak{sl}(n)}_{k=1}$ 
characters.

The $\widehat{\mathfrak{sl}(2)}_{k=1}$ case is re-examined first.
The novel formulation yields a direct connection to
the fractional statistics in the short range interacting model,
and provides a clear description of the spinon character formula.

Generalizing the observation, we find formulas for 
$\widehat{\mathfrak{sl}(n)}_{k=1}$,
which can be proved by the Durfee rectangle formula.
\end{abstract}

\noindent Short Title: Characters from TBA

\noindent 5.30.d, 11.25.Mf

\clearpage
\section{Introduction}

Quasi-Particles Character formulas have 
been an issue of current interests.
\cite{SUNY931,SUNY932,SUNY933,KNSch93,Metzer94,
Berkovich94, BerkMC96}
They are proposed in  various contexts; in the studies of 
lattice models, coset field theories or parafermionic field theories.
These formulae have backgrounds in physics;
they have originated from the description of
particle contents in terms of string solutions to
the Bethe Ansatz equations(BAE).

There may be, however, many ways to interpret and, 
thereby represent, affine characters in term of
quasi-particles.
We refer to studies on long range interacting models which 
provide a different point of view\cite{HHTBP,BPS,BLS94,
NakYam96,Yamada97,Nagoya}

Recently, a novel approach to describe
short range interacting 1D quantum systems at finite temperature has been
developed\cite{Suz85}- \cite{JutKluSuz97d}. 
It offers  different "basis" from strings, in a sense.
One might naturally expect a novel way to represent
affine characters in terms of these new basis.
 In this report, we will explicitly show that this expectation is indeed
realized  for affine level 1 $ \mathfrak{sl}(n)$
(hereafter referred to as $\widehat{ \mathfrak{sl}(n)}_{k=1}$) .

There could be several (quasi-)particles in a physical system.
Some of them describe the "Dirac sea", and
the others stand for excitations. 
The standard string approach does not distinguish
them, a priori.
Furthermore, the arbitrary lengths of strings are allowed for
the cases under consideration.\footnote{Our main interest concerns
with vertex models, not RSOS models in this report.}
Consequently, there exist infinite numbers of "(quasi-)particles", 
which makes the description of the affine characters difficult
in "string basis" (except for the case where
the numbers of (quasi-)particles reduce finite).
In the novel approach, the Dirac sea is decoupled
from others and the number of "particles" is finite.
Thus the affine character can be 
represented  by finite numbers of summation indices, which may
be associated to quantum numbers of "particles". 

This paper is organized as follows.
We will first  
re-visit the thermodynamics of the spin 1/2 XXX model, i.e., 
 the $\widehat{ \mathfrak{sl}(2)}_{k=1}$ case .
Some background in our approach will be reviewed, and
re-interpreted.
This simplest example already provides the essence of
our argument.
The novel formulation yields a direct
connection to the fractional statistics in this short range
interacting model.
In a sense, we are directly dealing with excitations which
reduces to spinon in $T \rightarrow 0$ limit.
We will see that the appropriate analytic continuation
of the Rogers dilogarithm function leads to the "spinon"
character for this case.
Generalizing the observation there, we will propose
a quasi-particle like character formula for 
$\widehat{ \mathfrak{sl}(n)}_{k=1}$  in section 3.
A brief summary and discussions are given in section 4.
%
%

\section{ $\widehat{ \mathfrak{sl}(2)}_{k=1}$  Revisited}

The thermodynamics of spin 1/2 XXX model has a long history of success.
The standard approach string hypothesis formulates it by the infinite numbers
of unknowns $\eta_m (m=1,\cdots,\infty)$ \cite{Gaudin, Tak71}. 
As noted in
the introduction, this is a drawback in presenting the
affine character or partition function.

A different approach is initiated in \cite{Suz85} and
subsequently progressed, especially in \cite{Klu93}.
It integrates two important ingredients,
the equivalence theorem between $D$ dimensional
quantum systems and $D+1$ dimensional classical systems, 
and the (Yang-Baxter) integrable structure.
 We start from a 
1D quantum system at $T=1/\beta$ with
system size $L$.
By introducing a fictitious dimension $M$, 
it can be mapped to a 2D classical vertex model defined on
a $L\times M$ square lattice.
We  first observe that 
the anisotropy in the coupling constant $u=-\beta/M $
"intertwines" finite temperature system to finite
size system.
The partition function can be represented in terms of
the "Quantum Transfer Matrix" (QTM) propagating in the
crossing channel.
The second observation is that QTM constitutes 
a commuting family labeled by a complex parameter $v$.

These observations can be summarized into the following formula for
the partition function by QTM,
$$
  {\cal Z}(\beta ) = \lim_{L\rightarrow \infty} \lim_{M\rightarrow \infty}
            {\hbox{ Tr }} (T_{\hbox{QTM}}(u=-\frac{\beta}{M}, v=0))^L.
$$

The existence of parameter $v$ actually
 leads to a  deeper result.
BAE for diagonalization of 
QTM can be recognized as the analyticity
condition of $T_{\hbox{QTM}}(u,v)$ in the complex
$v$ plane.
The investigation of the eigenvalue 
$T_{\hbox{QTM}}(u,v)$ then turns out to be that
of the analyticity of auxiliary functions in the complex $v$ plane.

The explicit eigenvalue $\Lambda(u,v)$ of $T_{\hbox{QTM}}(u,v)$
is composed of 2 terms $\lambda_1(u,v)$ and
$\lambda_2(u,v)$

\begin{equation}
\begin{split}
\Lambda(u,v) &= \lambda_1(u,v) + \lambda_2(u,v), \\
\lambda_1(u,v)&=\phi_-( v-i) \phi_+(v) \frac{Q_1(v+i)}{Q_1(v)} \qquad
\lambda_2(u,v)=\phi_+( v+i) \phi_-(v) \frac{Q_1(v-i)}{Q_1(v)} \\
\intertext{where}
\phi_{\pm} (v) &= (v\pm iu)^{M/2}, \qquad 
Q_1(v) =\prod_{j=1}^{m} (v-v_j)
\end{split}
\end{equation}
and $\{v_j \}$ is the solution to BAE,
$$
-1 = \frac{\phi_-(v_j-i) \phi_+(v_j)}{\phi_+(v_j+i) \phi_-(v_j)}
      \frac{Q_1(v_j+i)}{Q_1(v_j-i)}.
$$

We remark that each $\lambda_i$ carries spurious  
singularities, $1/Q_1(v)$. The BAE condition, however,
assures pole-freeness of the sum of them.
Respecting this the  auxiliary functions are introduced in \cite{Klu93},

\begin{equation}
\begin{split}
\mathfrak{a}^+(v)&= \frac{\lambda_1(u,v+i\gamma )}{\lambda_2(u,v+i\gamma)} 
  \qquad 
  \mathfrak{A}^+(v) = 1+\mathfrak{a}^+(v) \\
 \mathfrak{a}^-(v) &= \frac{\lambda_2(u,v-i\gamma)}{\lambda_1(u,v-i\gamma)} 
  \qquad 
 \mathfrak{A}^-(v) = 1+\mathfrak{a}^-(v) \\
\end{split}
\end{equation}
The shifts in arguments by $0<\gamma\le 1/2$ is introduced to avoid
zeros of $\mathfrak{A^{\pm}}$ on the real axis.

For the evaluation of the free energy, we only have to evaluate
 the largest eigenvalue of $T_{\hbox{QTM}}(u,v)$ due to
the existence of finite gap at finite temperatures.
Let us summarize the relevant results in \cite{Klu93, JutKluSuz97}.
The sector lies in $m=\frac{M}{2}$.
The auxiliary functions have the following strips 
where they are Analytic and Nonzero
and have Constant asymptotic values (ANZC);
\begin{eqnarray*}
\mathfrak{a}^+(v)\,\,  -u-\gamma \le \Im v \le 1+u-\gamma,
\qquad 
\mathfrak{A}^+(v)\,\, -\gamma< \Im v < 1-\gamma,  \\
\mathfrak{a}^-(v)\,\,  -1-u+\gamma \le \Im v \le u+\gamma,
\qquad 
\mathfrak{A}^-(v)\,\, -1+\gamma < \Im v < \gamma.
\end{eqnarray*}
Remark that $u$ is a negative small quantity.

Especially, the ANZC properties of
$ \frac{\mathfrak{a}^{\pm}(v-i\gamma)\phi_{\mp}(v)}{\phi_{\pm}(v)},
\frac{\mathfrak{A}^{\pm}(v-i\gamma)\phi_{\mp}(v)}{Q_1(v)}$
near the real axis leads to
 to  nonlinear integral equations,
 
\begin{equation}
\begin{split}
\log \mathfrak{a}^+(v) &=    -2 \pi \beta \Psi(v + i\gamma)
                           + K* \log \mathfrak{A}^+ (v) -
                           K* \log \mathfrak{A}^-(v+ 2i\gamma) \\
\log \mathfrak{a}^-(v) &=  2 \pi \beta \Psi(v-i\gamma)
                            + K*\log\mathfrak{A}^-(v) -
                           K* \log \mathfrak{A}^+(v- 2i\gamma)  \\
\intertext{ where }
\Psi(v) &= \frac{i}{2 \sinh \pi v}, 
\qquad K(v) = \frac{1}{2\pi}
              \int^{\infty}_{-\infty}  \frac{e^{ikv}}{1+e^{|k|}} {\rm d} k
\label{xxznlie}
\end{split}
\end{equation}
and $A*B(x)$ denotes convolution
 $\int^{\infty}_{-\infty} A(x-y) B(y) {\rm d }y$.

The free energy can be represented in an integral
form using $\log{\mathfrak{A}^{\pm}}$.
\begin{equation}
-\beta f = 2\beta \log 2 + 
    \int^{\infty}_{-\infty} 
 \Psi(v+i\gamma) * \log \mathfrak{A}^+(v) {\rm d} v -
  \int^{\infty}_{-\infty}  \Psi(v-i\gamma) * \log \mathfrak{A}^-(v) {\rm d} v .
\label{xxxfree}
\end{equation}

The first term in the lhs is the well-known ground state energy.
We thus  manifestly separate the contribution by the Dirac sea.  
Thermal fluctuations are described by $\mathfrak{a}^{\pm}$.
Although the integrands in (\ref{xxxfree}) have fictitious dependence on 
$\gamma$, the result of integration is independent of
its value.
This comes from the 
ANZC properties of $\mathfrak{A}^{+}$ and $\mathfrak{A}^{-}$ 
within strips described above. Note $\gamma\le 1/2$.

The relevant result for the present discussion is the following
expression for the central charge.

\begin{equation}
c_0= L(\frac{\mathfrak{a}^+(\infty)}{\mathfrak{A}^+(\infty)}) +
 L(\frac{\mathfrak{a}^-(\infty)}{\mathfrak{A}^-(\infty)})  \qquad (=1)
\end{equation}
where $L(z)$ is Rogers dilogarithm function
\begin{equation}
L(z) = -\frac{1}{2} \int^{z}_{0} \Bigl (\frac{\log(1-x)}{x}+ 
  \frac{\log(x)}{1-x} \Bigr)
  {\rm d}x
\label{dilog}
\end{equation}

which has branch cut along lines $(-\infty,0], [1,\infty )$.
The partition function can be, after subtraction of the
ground state energy, approximated to
$$
{\cal Z}(\beta ) \sim  q^{-c_0/24}
$$
where $q= \exp(-4\pi/ v_F \beta)$ and $v_F$ is the Fermi velocity.

Now we reinterpret these results.
First, we note a result from numerical investigation.
Fix $\gamma=1/2$. 
For sufficiently low temperatures, 
one observes a  crossover behavior of $\mathfrak{a}$ 
in very narrow regions ($ v \sim \pm \log \beta $).
This justifies the approximation of 
the kernel function $K$ in 
(\ref{xxznlie} ) to $\frac{1}{2} \delta(v)$.
Assuming the validity of this approximation, 
we perform the change of variable $k+\pi/2 =2 \tan^{-1}(e^v)$.
By writing $ w_1(k)=1/\mathfrak{a}^+,w_2(k)=1/\mathfrak{a}^-$,
the resultant equations now can be casted into the form,

\begin{equation}
\begin{split}
\exp(\beta \epsilon_{a}^0(k)) &=(1+w_{a})
   \prod_{b=1,2} \Bigl (\frac{w_{b}}{1+w_{b}} \Bigr ) ^{\mathfrak{g}_{b,a}}  \\
f &= -T \sum_{a=1,2} \int G_a \log(1+\frac{1}{w_a}) d k  \\
\epsilon_{a}^0(k)&= \cos k, \qquad G_1 = G_2 =\frac{1}{2} .
\label{fss}
\end{split}
\end{equation}

This is nothing but the equation describing particles
obeying a 
fractional exclusion statistics \cite{Haldane91,BerWu94}
, where 
statistical interaction matrix is given by
\begin{equation}
\mathfrak{g} = \begin{pmatrix}
             1/2,& 1/2& \\
             1/2,& 1/2& 
              \end{pmatrix}  
\label{gforsl2}
\end{equation}
This coincides with known $\mathfrak{g}$ for spinons \cite{Schoutens97}.
This is an explicit  demonstration of
the fractional statistics in the solvable spin chain model,
although in an approximate sense.
One recognizes a clear difference between the above result and
the fractional exclusion statistics approach to the same model 
based on the conventional string picture.
See the appendix of \cite{KatoKura}.

We further exploit the coincidence.
Consider the conformal limit (or character limit).
The higher terms are now contributing to ${\cal Z}$.
The analysis of contributions from excited states 
is of considerable interest.
See \cite{KPexc}-\cite{PZ} for progress in related areas
and \cite{Klu93,JutKluSuz97c, KSS} in the context of QTM.
For $T$ finite, it requires the explicit evaluation 
of zeros of auxiliary functions
entering into "physical strip" which is free from zeros
and singularities for the largest eigenvalue sector.
This involves extensive numerical studies.
See \cite{Klu93, BLZ, PF, KSS} for example.
Since our concern here is the asymptotic behaviors,
we can avoid such numerics by
following the strategy in \cite{KNSch93}.
By adopting general contour ${\cal C}$ for the integration in
(\ref{dilog}),
one can define an analytically continued dilogarithm function
$L_{\cal C}(z)$.
Such contour dependent dilogarithm function successfully
recovered conformal spectra for lower excitations.
We assume this in general;
 all excitation spectra in the conformal limit shall be
described by analytically continued dilogarithm function.

Let ${\cal C}$ be a contour starting from origin and 
terminating at $f$, such that it 
crosses firstly $[1, \infty)$ $\eta_1$ times, then 
crosses  $(-\infty,-1]$ $\xi_1$ times, then
again $\eta_2$ times and so on.
We adopt the convention $\eta_j, \xi_j <0$ if the contour
crosses branch cuts clockwise and $\eta_j, \xi_j >0$
otherwise.
Let us denote this by 
${\cal C}[f|\xi_1, \xi_2, \cdots |\eta_1, \eta_2, \cdots]$.
Then  contours are  parameterized by a set $\{ {\cal S} \}
=\{ {\cal C}^{\pm}[f^{\pm}| \xi^{\pm}_m | \eta^{\pm}_m] \}$, where
$f^{\pm} = \mathfrak{a}^{\pm}(\infty)/\mathfrak{A}^{\pm}(\infty)$
The effective central charge can be expressed by
\begin{equation}
\begin{split}  
  c({\cal S}) &= c_0 - 24 {\cal T}({\cal S}) ,  \qquad
  {\cal T}({\cal S}) =
  \frac{1}{2} \phantom{}^t \mathbf{n} \mathfrak{g}\mathbf{n} -
          \sum_{a=\pm} \sum_{j=1, \cdots} \xi^{a}_j 
                        (\eta^{a}_1+\cdots +\eta^{a}_j)  \\
\mathbf{n} &=   \begin{pmatrix}
                    n^+\\
                    n^-\\
                   \end{pmatrix} 
\end{split}
\end{equation}
where $  n^{\pm}  = \sum_{j} \eta^{\pm}_j.$

Following \cite{KNSch93}, we choose a subset ${\cal O}$
of all possible contours,
$$
\{ {\cal C}^{\pm}[f^{\pm}| \xi^{\pm}_1, \cdots, \xi^{\pm}_{n^{\pm}} | 
 \overbrace{1,1,\cdots, 1}^{n^{\pm}}, 0, \cdots, ] \}
$$
and $\xi^{\pm}_j \le 0, j=1, \cdots, n^{\pm}$.

The summation over ${\cal O}$ leads to the following
"partition function".
\begin{equation}
 \sum_{{\cal S} \in {\cal O}} q^{-c({\cal S})/24} =  
  q^{-c_0 /24} \sum_{n^+,n^-}  
   \frac{q^{1/2  \phantom{}^t \mathbf{n} \mathfrak{g}\mathbf{n}       }}
        {(q)_{n^+} (q)_{n^-}}
   \label{spinonch}
 \end{equation}
where $(q)_m= \prod_{i=1}^m (1-q^i)$.
This is nothing but the spinon character formula.

The explicit forms of affine characters have
been obtained in mathematical literatures,
e.g.,  \cite{KacPeter}.
Let $ \alpha$ be the simple
for $\mathfrak{sl}_2$ and $Q=\mathbf{Z} \alpha$.
Then $\widehat{\mathfrak{sl}(2)}_{k=1}$ character reads,

\begin{equation}
\hbox{ch } L(\Lambda) (q,u) = \frac{1}{\eta(q)} 
 \sum_{\alpha \in Q} q^{(\alpha+\bar{\Lambda})^2/2}
   e^{2\pi i <\alpha+\bar{\Lambda}, u>}
\label{chsl2}
\end{equation}
where $\bar{\Lambda}$ denotes the classical part
and $\eta(q) = q^{1/24} \prod_{j\ge 1} (1-q^j)$.
We use the notation $z= e^{2\pi i <\alpha,u>}$.

Our partition function may trace all energy levels and 
their degeneracies. Thus it should
coincide with (\ref{chsl2}) after appropriate modification
due to  $z$.
Actually the following theorem holds \cite{SUNY933,BLS94, FeiginStoy},

\begin{theo}
The character for $\widehat{\mathfrak{sl}(2)}_{k=1}$
assumes the following Fermionic form,

$$
\hbox{ch } L(\Lambda) (q,z) =
  q^{-c_0 /24} \sum_{n^+,n^-} 
   \frac{q^{1/2 \mathbf{n}\mathfrak{g}\mathbf{n}  }}{(q)_{n^+} (q)_{n^-}}
   z^{1/2(n^+ -n^-)}
$$
where  $n^+ +n^- =$  even (odd) for $\Lambda=\Lambda_0 (\Lambda_1)$.
\end{theo}

The Fermionic character  further implies 
that our auxiliary functions indeed describe 
the spinons \cite{JKM73, FadTak81}.

Let us summarize the lesson from  $\widehat{\mathfrak{sl}(2)}_{k=1}$.
We define auxiliary functions respecting the cancellation of
spurious poles. 
These auxiliary functions  satisfy coupled integral equations.
In the "character limit", the partition function is
described by $\mathfrak{g}$ which can be
readily evaluated form asymptotic behaviors of the coupled integral equations.
Taking account of winding numbers of dilogarithm function, we
reach the quasi-particle character formula.
%
%

\section{ Novel Character Formula for 
$\widehat{ \mathfrak{sl}(n)}_{k=1}$   }

Next we consider  
$\widehat{ \mathfrak{sl}(n)}_{k=1}$  $n=r+1$ arbitrary. 

The eigenvalue of QTM consists of $r+1$ terms,
\begin{equation}
\begin{split}
\lambda_a(v) &=\phi_a(v)
  \frac{Q^{(a-1)}(v-i) Q^{(a)}(v+i)} {Q^{(a-1)}(v) Q^{(a)}(v)}, 
(a=1,\cdots,r+1)\\
 \phi_a(v) &=\begin{cases}
                   a=1,&    \phi_+(v) \phi_-(v-i) \\
                   a=r+1,&  \phi_+(v+i) \phi_-(v) \\
                   {\hbox{otherwise }}&  \phi_+(v) \phi_-(v)
             \end{cases}
\qquad
Q^{(a)}(v)= \begin{cases} 
                   a=0 {\hbox{ or }} r+1,&   1\\
                   {\hbox{otherwise }}& \prod_j(v-v^{(a)}_j).
             \end{cases}
\end{split}
\end{equation}
Each neighboring terms share common denominators
which seem to bring singularity.
BAE elucidates these fictitious poles,
$$
{\hbox{Res}}_{v= v^{(a)}_j} (\lambda_{a}(v)+\lambda_{a+1}(v))=0.
$$
(For cancellations in the context of analytic Bethe ansatz, 
see \cite{KSanaly95} .)

More generally, in the
 string of terms $\lambda_{r+1}+\cdots+\lambda_{i}$,  
most of their "singularities" are canceled out 
but ones from $\lambda_{i}$.
These extra poles  can be canceled by adding $\lambda_{i-1}$ which 
brings about other singularities.
The proper choice of the auxiliary functions may respect this
"pole canceling" property,
\begin{equation}
\begin{split}
\mathfrak{a}^+_i(v) =& 
 \frac{\lambda_{r-i+1}(v+i\gamma)}{(\lambda_{r+1}(v+i\gamma)+
  \lambda_{r}(v+i\gamma) + \cdots \lambda_{r-i+2}(v+i\gamma))},
                          \qquad 
\mathfrak{A}^+_i(v)=1+\mathfrak{a}^+_i(v) , (i=1, \cdots,r) \\
\mathfrak{a}^-_i(v) =& 
 \frac{(\lambda_{r+1}(v-i\gamma)+\lambda_{r}(v-i\gamma) +
                          \cdots \lambda_{r-i+2}(v-i\gamma))}
	  {\lambda_{r-i+1}(v-i\gamma)},
                          \qquad 
\mathfrak{A}^-_i(v)=1+\mathfrak{a}^-_i(v) , (i=1, \cdots,r).
\end{split}
\end{equation}

These choices actually generalize the known results for 
$r=1$ in the previous section and $r=2$ \cite{KWZ97}.
\footnote{ I thank Andreas Kl{\"u}mper for the discussion on
this point.}
The straightforward generalization of the ANZC type
arguments leads to the following equation for the asymptotic
values
$$
\widetilde{\log} \mathfrak{a}(\infty) = 
{\cal K} \widetilde{\log} \mathfrak{A} (\infty)
$$
where $\widetilde{\log} \mathfrak{a} $ is the abbreviation of the
column vector, $( \log \mathfrak{a}^+_1, \cdots,
 \log \mathfrak{a}^+_r, \log \mathfrak{a}^-_1, \cdots,
 \log \mathfrak{a}^-_r)$ and similarly for $ \widetilde{\log} \mathfrak{A} $.
The generalization  of $\mathfrak{g}$ in (\ref{gforsl2}) is
now explicitly given by

$$
\mathfrak{g} = I-{\cal K} =
   \begin{pmatrix}
    C^{-1}_{\mathfrak{sl}_{r+1}},&  C^{-1}_{\mathfrak{sl}_{r+1}}(I-U)\\
    (I-D) C^{-1}_{\mathfrak{sl}_{r+1}},&
     (D-I) C^{-1}_{\mathfrak{sl}_{r+1}}(U-I) \\
             \end{pmatrix}
$$
where $C_{\mathfrak{sl}_{r+1}}, I$ denotes the Cartan matrix
for $\mathfrak{sl}_{r+1}$ and the $r\times r$ identity matrix,
respectively. $D$ and $U$ is defined by
$D_{i,j} = \delta_{i,j+1}, U=\phantom{}^t D$.
Repeating similar arguments as in the case of 
$\widehat{\mathfrak{sl}(2)}_{k=1}$,
we have the partition function,

\begin{equation}
\begin{split}
 {\cal Z} & = q^{-r/24} 
 \sum_{n^{\pm}_j \ge 0} 
  \frac{q^{\phantom{}^t \mathbf{n} \mathfrak{g}\mathbf{n}/2}}
       {(q)_{\mathbf{n}}}   \\
\intertext{ where }
\phantom{}^t \mathbf{n} &=  (n^+_1, \cdots, n^+_r, n^-_1, \cdots, n^-_r),
\qquad
(q)_{\mathbf{n}} = \prod_{j=1}^r (q)_{n^+_j} (q)_{n^-_j}.
\end{split}
\end{equation}

Affine character for $\widehat{\mathfrak{sl}(r+1)}_{k=1}$ 
which generalizes eq.(\ref{chsl2}) is given by 
\begin{equation}
\hbox{ch } L(\Lambda) (q,z_1, \cdots, z_r) = \frac{1}{\eta(q)^r} 
 \sum_{\alpha \in Q} q^{(\alpha+\bar{\Lambda})^2/2}
   e^{2\pi i <\alpha+\bar{\Lambda}, u>}
\label{chsln}
\end{equation}
where  $Q= \sum_{i=1}^{ r} \mathbf{Z} \alpha_i, z_i= e^{2\pi i <\alpha_i, u>}$,
 $z^{\mathbf{ n}} = \prod_{j=1}^r z_j^{n_j}$.

Let us present the main result in this report.

\begin{theo}
Let $L(\Lambda_a)$ be the highest weight module for affine
Lie algebra $\widehat{\mathfrak{sl}(r+1)}_{k=1}$ with the highest
weight $\Lambda_a$.
Then the character has a quasi-particle like form,
\begin{equation}
{\hbox{ch}}L(\Lambda_a) (q,z)  = 
 q^{-r/24} \sum_{n^{\pm}_j \ge 0} 
  \frac{q^{\phantom{}^t \mathbf{n} \mathfrak{g}\mathbf{n}/2}}
       {(q)_{\mathbf{n}}} 
        z^{\mathbf{\mu}(\mathbf{n})}  
\label{quasisln}
\end{equation}
 where  the exponent of $z$ reads
\begin{equation}
\begin{split}
\mathbf{\mu}(\mathbf{n}) &=
   \phantom{}^t (\mathfrak{W}^{-1}) 
    C^{-1}_{\mathfrak{sl}_{r+1}} 
	((I-D)^{-1} \bf{n}^{+} -\bf{n}^{-} ),
 \qquad
 \mathfrak{W}=C_{\mathfrak{sl}_{r+1}} \mathfrak{W}^\vee 
              C^{-1}_{\mathfrak{sl}_{r+1}}, \\ 
\mathfrak{W}^\vee  &= S*(I-U)^{-1}, \qquad
(S*A)_{i,j} = A_{i,\tilde{j}}, \quad  \tilde{j}=j+1 ( \hbox{Mod } r).
\end{split}
\end{equation}

The summation is subject to the rule;
$ (\mathfrak{g} \mathbf{n})_1=1-\frac{a}{r+1} (\hbox{Mod } \mathbf{Z})$.

\end{theo}

For the proof of the above formula, we prepare some lemmas.

\begin{lemm}["Durfee rectangle formula"]
$$
\frac{1}{(q)_{\infty}}= \sum_{a,b\ge 0, a-b=m} \frac{q^{ab}}{(q)_a (q)_b}
$$
Here the difference $m$ is an arbitrary but fixed integer.
\label{Durfee}
\end{lemm}

We prove the equivalence between eq.(\ref{chsln}) and 
eq.(\ref{quasisln}) by 
fixing weight $\mu(\bf{n})$.
This allows us to rewrite equations in term of $\{n^-_j \}$,
as it imposes relations between $\{n^+_j \}$ and $\{n^-_j \}$;
\begin{equation}
\bf{n}^+ = (I-D) (\bf{n}^- +\nu) 
\qquad
\hbox{where   } \,\,
\nu=C_{\mathfrak{sl}_{r+1}} \phantom{}^t \mathfrak{W} \mu(\bf{n}).
\label{relationpm}
\end{equation}


\begin{lemm}

Under eq.(\ref{relationpm}), the q-exponent in eq.(\ref{quasisln}) can be 
expressed as 
\begin{equation}
\begin{split}
\phantom{}^t \mathbf{n} \mathfrak{g}\mathbf{n}/2
   &= {\cal{F}}(\nu) +{\cal {G}}(\nu, n^{-}) ,   \\
{\cal{F}}(\nu)&=
  \phantom{}^t \mathbf{\nu} (I-U) C_{\mathfrak{sl}_{r+1}}^{-1} 
  (I-D)\mathbf{\nu}/2
  \qquad
{\cal{G}}(\nu)=\frac{1}{2}(
  \phantom{}^t \mathbf{\nu} (I-U) \mathbf{n}^- +
   \phantom{}^t \mathbf{n}^- (I-D) \mathbf{\nu} +
   \phantom{}^t \mathbf{n}^-  C_{\mathfrak{sl}_{r+1}}   \mathbf{n}^- )
\end{split}
\end{equation}
\end{lemm}
This can be easily derived with aid of the trivial relation, 
\begin{equation}
C_{\mathfrak{sl}_{r+1}} = (I-U) +(I-D).
\label{trivial}
\end{equation}

Summation condition is now simplified and  independent of $n^-$. 
To be precise, the following lemma holds.

\begin{lemm}

The summation condition, 
$ (\mathfrak{g} \mathbf{n})_1=1-\frac{a}{r+1} (\hbox{Mod } \mathbf{Z})$
is equivalent to 
$\nu_1+\cdots+\nu_r= r+1-a (\hbox{Mod } r+1)$.
The latter condition is indeed satisfied by eq(\ref{relationpm}).
\label{sumcond}
\end{lemm}

A straightforward calculation shows that the original summation condition 
reduces to
$n_1-1=(a+\nu_1+\cdots+\nu_r)/r+1 (\hbox{Mod } \mathbf{Z})$, which proves
the former part of the lemma.
The latter part can be also easily shown by
noticing that $\alpha$ in eq.(\ref{chsln}) is a member of $Q$.
Let $\alpha= \sum m_i \alpha_i, m_i \in \mathbf{Z}$.
Equating the exponent of $z$ in eqs.(\ref{chsln}), 
(\ref{quasisln}),
we have $\nu= {\bf{m}}+ C_{\mathfrak{sl}_{r+1}}^{-1} {\bf{e}}_a$, where
$e_a$ denotes the unit vector with only its $a-$ th entry unity and all others
zero. By substituting this into eq.(\ref{relationpm}), 
we find $\nu_1+\cdots+\nu_r=(r+1) m_1+r+1-a$. 
Thus the latter part is proved.
\begin{lemm}

In terms of $\{ m_i \}$ defined in the above, we have 
\begin{equation}
{\cal{F}}= \phantom{}^t {\bf{m}} C_{\mathfrak{sl}_{r+1}} {\bf{m}} +
           2 m_a +(C_{\mathfrak{sl}_{r+1}}^{-1})_{a,a}.
   \end{equation}
\label{finalF}
\end{lemm}
This is a consequence of the relation,
$$
C_{\mathfrak{sl}_{r+1}}^{-1}=
 \mathfrak{W}^{\vee} (I-U) C_{\mathfrak{sl}_{r+1}}^{-1}
(I-D) \phantom{}^t(\mathfrak{W}^{\vee}) ,
$$
which can be shown by a direct calculation.

\noindent{\it{Proof of Theorem 2 :}} 
We first fix $\mu$.
Due to Lemma 3,4, for each fixed $\mu$, eq.(\ref{quasisln})
 is now rewritten as
$$
q^{-r/24} z^{\mu} q^{\cal{F}} \sum_{n^-_j} 
 \frac{q^{\cal{G}}} {(q)_{\bf{n}}}
$$
where $\{n^+_j \}$ in the denominator are functions of $\{n^-_j\}$
by eq(\ref{relationpm}).
Thanks to Lemma 4, the summations over $\{n^-_j\}$
are almost non-restricted except for
" implicit " conditions, $n^+_j \ge 0$.
This actually meets the condition of  Lemma 1.
Let us see this explicitly for the summation over $n^-_1$.
Note first that ${\cal {G}} = (n^-_1)^2 -n^-_1 n^-_2 
+ n^-_1(\nu_1-\nu_2) +$ (terms independent of $n^-_1$), 
and $n^+_1 = n^-_1-n^-_2+\nu_1-\nu_2$.
Thus the contribution from the $n^-_1$ summation reduces to
$$
\sum_{n^-_1 \ge n_1^*} 
\frac{q^{(n^-_1)^2 -n^-_1 n^-_2+n^-_1(\nu_1-\nu_2)}}
                   {(q)_{n^-_1} (q)_{n^-_1-n^-_2+\nu_1-\nu_2} }
$$
where $n_1^* :=  \hbox{max}(0, n^-_2+\nu_2 -\nu_1) $.
Identifying $a=n^-_1, b=n^-_1-n^-_2+\nu_1-\nu_2, m= -n^-_2+\nu_1-\nu_2$,
we can apply Lemma 1 straightforwardly.
Thus the contribution from 
$n^-_1$ part is simply given by $1/(q)_{\infty}$.

Repeating this successively we reach the expression,
$
\frac{ z^{\mu} q^{\cal{F}}}{\eta(q)^r}.
$
By Lemma \ref{finalF}, this can be  entirely written in term of
$\{m \}$ as,
$$
\frac{z^{ {\bf{m}} + C_{\mathfrak{sl}_{r+1}}^{-1} {\bf{e}}_a } 
      q^{ \phantom{}^t {\bf{m}} C_{\mathfrak{sl}_{r+1}} {\bf{m}} +
           2 m_a +(C_{\mathfrak{sl}_{r+1}}^{-1})_{a,a} }
	}{\eta(q)^r},
$$
which agrees with a term in eq(\ref{chsln}) with $\alpha=\sum m_i \alpha_i$
and $\Lambda=\Lambda_a$.
Thus the proof is completed. $\square$

As demonstrated in $\widehat{\mathfrak{sl}(2)}_{k=1}$, 
the Dirac sea contribution is separated and 
$\{ \mathfrak{a}^{\pm}_i \}$ correspond to thermal excitations.
Thus we may expect that they describe physical particles.

\section{Summary and Discussion}

In this report, we present a  character formula of
the Fermionic type for $\widehat{\mathfrak{sl}(r+1)}_{k=1}$.
It has originated from the novel formulation of thermodynamics
of 1D quantum systems at finite temperatures.

The interpretation of the resultant formula in terms of 
particles, however, poses problems.
By the coincidence for $r=1$ case, we expect
relations to  $su(r+1)$ spinons \cite{BouwkSchou96}.
Unfortunately, this seems to be false.
For an example, 
the $su(3)$ spinon character needs 6 summation variables
(or 3)\cite{BouwkSchou96,NakYam96} in 
contrast to 4 variables in the present result.

We may assume there exist  "particles" $k^{(\pm)}$ corresponding 
"quantum numbers" $n_k^{\pm}$.

From (\ref{quasisln}),
we identify the weights of them as in the table \ref{tab1}

\begin{table}[htbp]
\begin{center}
\begin{tabular}{cc}
     "Particle" &                  Classical Weight   \\  \hline
   $1^{(+)}$  &                 $\Lambda_1$  \\
  $a^{(+)}, (2\le a \le r-1)$ & $  \Lambda_{a+1} -\Lambda_1$ \\
   $r^{(+)}$ &                  $\Lambda_1-\Lambda_2    $\\ 
%
   $a^{(-)}, (1\le a \le r-1)$ & $\Lambda_{a+2}-\Lambda_{a+1}$ \\
   $r^{(-)}$ &                  $\Lambda_2-\Lambda_1 $\\  \hline
%
\end{tabular}
\label{tab1}
\end{center}
\end{table}

"Particle" $a^{(+)}$ carries the weight in the highest weight module 
$V(\Lambda_a)$, while $a^{(-)}$ has the one in $V(\Lambda_1)$.
In a sense, $\pm$ "particles" are quite asymmetric, which does not seem
to be natural.
The vertex operators associated to negative or positive roots
bring a similar but "symmetric" $q-$character 
for whole vacuum module \cite{Georgiev}
\begin{equation}
\sum_{n^{\pm}_1, \cdots, n^{\pm}_r \ge 0}
\frac{ q^{ \sum_{j,k} 
           (n^+_j -n^-_j) ( C^{-1}_{\mathfrak{sl}_{r+1}})_{j,k} (n^+_k -n^-_k)
		   + \sum_{j} n^+_j n^-_j}}   {(q)_{\mathbf{n}}} 
\label{geor}
\end{equation}
under  $r$  constraints, 
$\sum_k  (C^{-1}_{\mathfrak{sl}_{r+1}})_{j,k} (n^+_k -n^-_k) \in \mathbf{Z},
 (j=1, \cdots, r)$. 
This formula also follows from a specialization of eq.(\ref{chsln})
using lemma 1.
This similarity makes us expect that
vertex operator description of  particles in eq.(\ref{quasisln}) 
would provide an interesting future problem. 

Obviously, several generalization awaits us, such as
the extension to  higher levels or to other affine
Lie algebras.
The higher level cases correspond to fusion models
in the terminology of lattice models.
%
%
The problem which shares the technical similarity to the present
one, the finite size correction study,
 has been successfully investigated  {\it only} up to level 2
$\widehat{\mathfrak{sl}(2)}_{k=2}$.
The result there suggests more elaborated
combinations of $\lambda_i$ are necessary for auxiliary
functions \cite{KlumB90,KlumBP91}. 
 For general cases, the strategy
for suitable choices of auxiliary functions are yet to be known.
We hope that the present result triggers further investigations
on these general cases, which should be
properly interesting in view of the original problem of 
thermodynamics. 

\section*{Acknowledgments}

The author would like to thank A. Berkovich, B. McCoy
and Y. Yamada for discussions.
Special thanks are due to A. Kl{\"u}mper for discussions on
 the choice of auxiliary functions and 
 A. Kuniba for the essential idea in the
proof of the main theorem in this report.

\end{document}